\documentclass[aps,prc,twocolumn,floatfix,superscriptaddress]{revtex4}
\usepackage{epsfig}
\usepackage{amsmath}
\usepackage{color}
\usepackage[utf8]{inputenc}
\usepackage{graphicx}

\definecolor{blue}{rgb}{0.05, 0.05, 0.5}

\begin{document}

\title{Bottom Production in $pp$ Collisions at Large Hadron Collider Energies using Parton Cascade Model}
\author{Rupa Chatterjee}
\email{rupa@vecc.gov.in}
\author{Dinesh K. Srivastava}
\email{dinesh@vecc.gov.in}
\affiliation{Variable Energy Cyclotron Centre, HBNI, 1/AF, Bidhan Nagar, Kolkata 700064, India}

\begin{abstract}
We study the production of bottom quarks in $pp$ collisions at Large Hadron Collider energies using
 previously developed parton cascade model to explore the impact of Landau Pomeranchuk Midgal (LPM)
 effect on their dynamics. In contrast to
the case for charm quarks reported recently, we find only a marginal impact of the suppression of
 multiple scatterings of partons due to the LPM effect on their production. 
It is felt that this happens as they are only produced in very hard collisions.  
 \end{abstract}


\maketitle

\maketitle
\section{Introduction}
The study of bottom production in relativistic collisions of protons as well as heavy nuclei
is quite important for a variety of reasons. Their production in $pp$ collisions is expected to provide
a stringent test of perturbative Quantum Chromo Dynamics as the $Q^2$ for their production 
$(\geq 4m_b^2)$ is of the order of 64--100 GeV$^2$. Their decay provides an important contribution
 to charm and $J/\Psi$ production. An accurate description of their production is an important prerequisite
in a search for new physics. It is also expected that the angular correlation of $b$ and $\overline{b}$
may provide insight into the influences affecting their movements in the plasma as well as, into the mechanism 
of their production.

Their reasonably abundant production in $AA$ collisions at higher energies, 
the recent and planned increase in the beam-luminosity and 
the detector systems for the experiments
at Large Hadron Collider (LHC) elevate them to a unique position in the  
study of relativistic heavy ion collisions, which are being investigated in order to explore
the dynamics of production of Quark Gluon Plasma (QGP) and its properties. The
QGP is predicted to exist 
by lattice QCD calculations (see Refs.~\cite{Ratti:2016lrh,Ratti:2017qgq,Ratti:2018ksb}
 and references therein). It is also believed that QGP filled the nascent Universe 
till about a few microseconds after the Big Bang. 

As bottom quarks are produced 
very early in the collision ($\tau\approx 1/2m_b$), they, or rather
 the $b\overline{b}$ duo, will be a witness to the
evolution of the system of
initial partons- from a pre-equilibrium stage of energetic partons to a thermalized
and possibly chemically equilibrated QGP due to a vehement multiplication of partons and
their scatterings and
then to a hadronised state, before undergoing freeze-out to a system of hadrons. 
In the mean time the $b\overline{b}$ would have either formed an $\Upsilon$ (or one of 
its excited states)
or the bottom and anti-bottom quarks would have drifted apart, buffeted by the 
energetic quarks and gluons and weakening
of the colour force between them due to Debye screening, to be hadronized to form B-mesons. 

Do bottom quarks thermalize
in the plasma? Are they affected by the hydrodynamic flow which is believed to develop in the system? If yes,
to what extent?  Do they
lose energy in the plasma (see e.g., 
\cite{Svetitsky:1987gq,GolamMustafa:1997id,Mustafa:1997pm,Djordjevic:2003zk,Djordjevic:2006tw,Abir:2012pu},
and references therein) due to collision with partons or radiation of gluons or both? Can they be developed
as an effective tool to determine the flavour dependence of jet quenching? These and many
other questions are expected to be answered in precise detail in near future. 
The usual first step in this direction is a
comparison of their production in nucleus-nucleus collisions with those for appropriately normalized (by
number of collisions) $pp$ collisions.

This procedure has come under strain due to recent findings which found "QGP like" features in $pp$
collisions, especially in events having large multiplicities~\cite{ALICE:2017jyt,Khachatryan:2016txc}. 
A recent theoretical study of $pp$ collisions at LHC 
energies~\cite{Srivastava:2018dye} within Parton Cascade Model (PCM)
suggested onset of a substantial multiple scattering, necessary for the formation
of an interacting system. A further support for this was provided by the study of charm production
which gave indications of  deviations in results for calculations performed with and without multiple
 scatterings, already in minimum bias events for $pp$ collisions at LHC energies~\cite{Srivastava:2018nfu}.
The experimental data were found to indicate a preference for results of the calculations obtained with the 
inclusion of Landau Pomeranchuk Midgal (LPM) effect
${\it and}$ multiple scatterings for low transverse momenta and central rapidities.

Does this happen for bottom production as well? Considering that bottom quark mass is much larger
than the charm quark mass, we expect that they may be produced only in the initial hard scatterings
and that the subsequent (multiple) scatterings may not entail a sufficiently large momentum transfer for their
production. The same consideration would also affect the $g \rightarrow b\overline{b}$ fragmentation, as it would involve a very large virtuality for the scattered gluon, in our description.
Thus, even if there is an increased mutiple scattering in (high multiplicity)
$pp$ collisions, the bottom production may differ only marginally for results of calculations
with and without inclusion of multiple collisions, once the LPM effect is accounted for.

 If confirmed, this would provide a justification to continue to
use the so-called medium modification factor "$R_{\text {AA}}$" with confidence 
as an accurate measure of the medium modification
of the bottom production due to the formation of QGP in $AA$ systems. We put this to test in the
following and find that it is indeed so.  

We very briefly discuss the necessary formulation in the next scion, followed by a section giving
our results. Finally we give our conclusions.

\begin{figure}[h]
\centerline{\includegraphics*[width=7.6 cm]{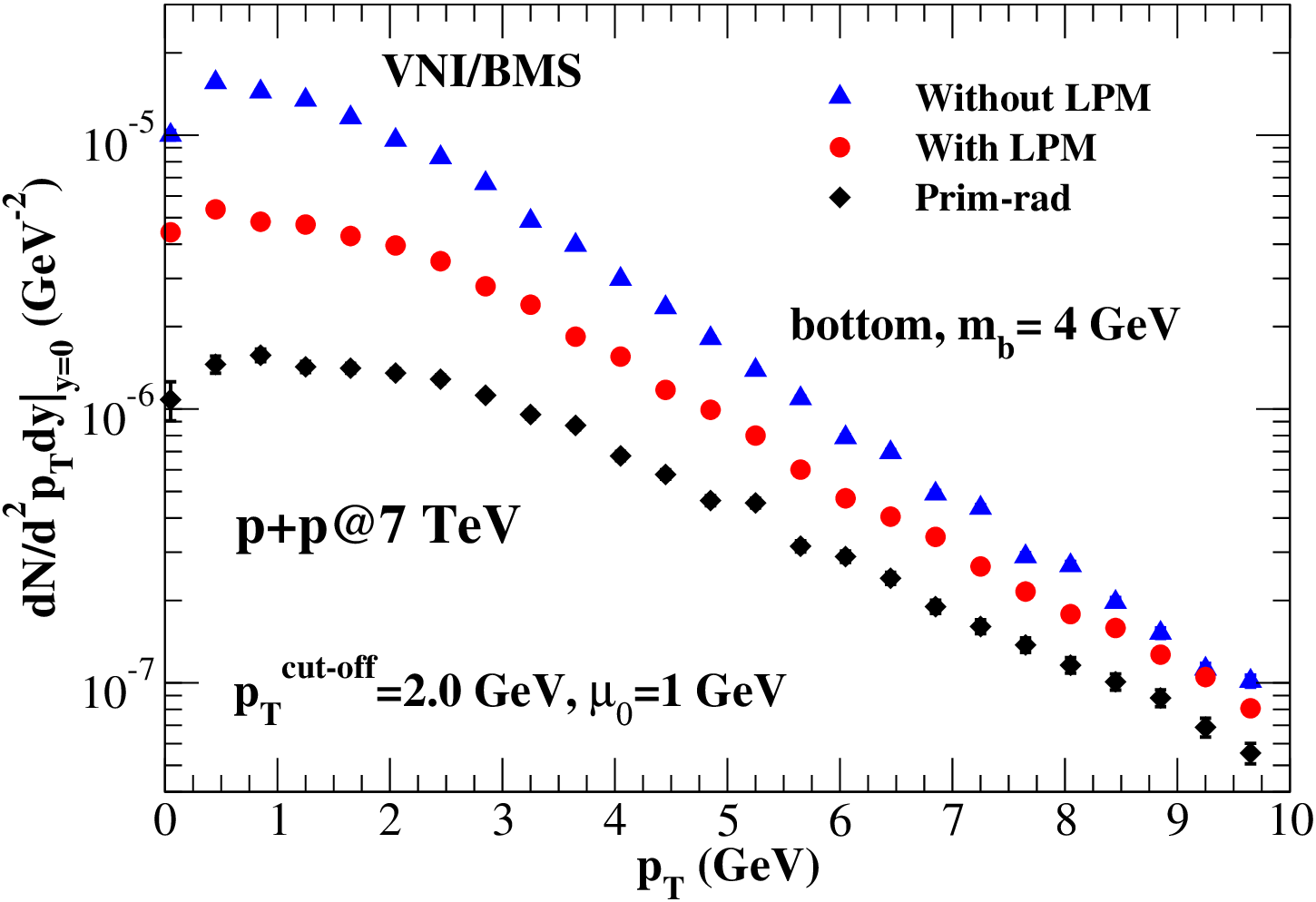}}
\centerline{\includegraphics*[width=7.6 cm]{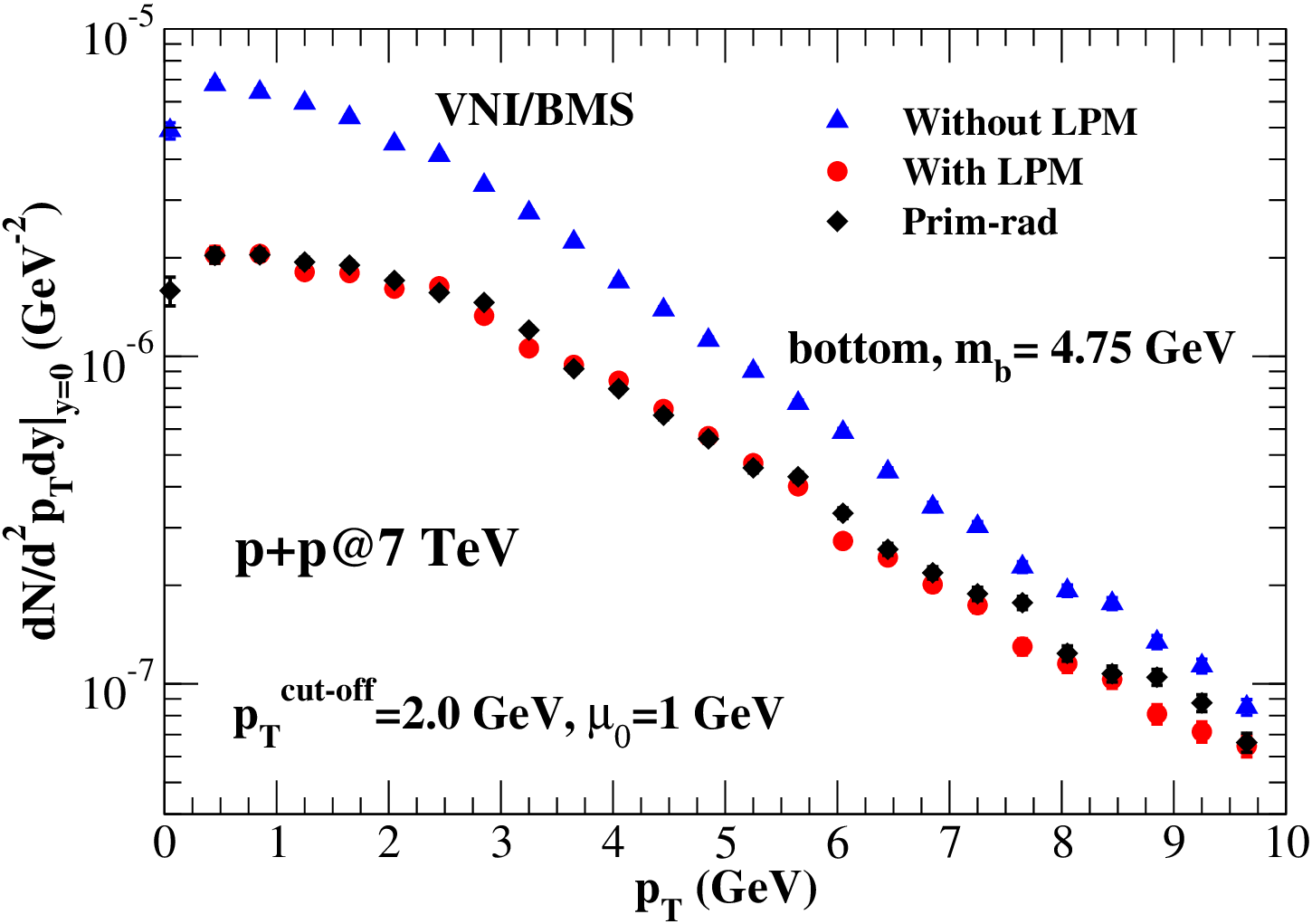}}
\centerline{\includegraphics*[width=7.6 cm]{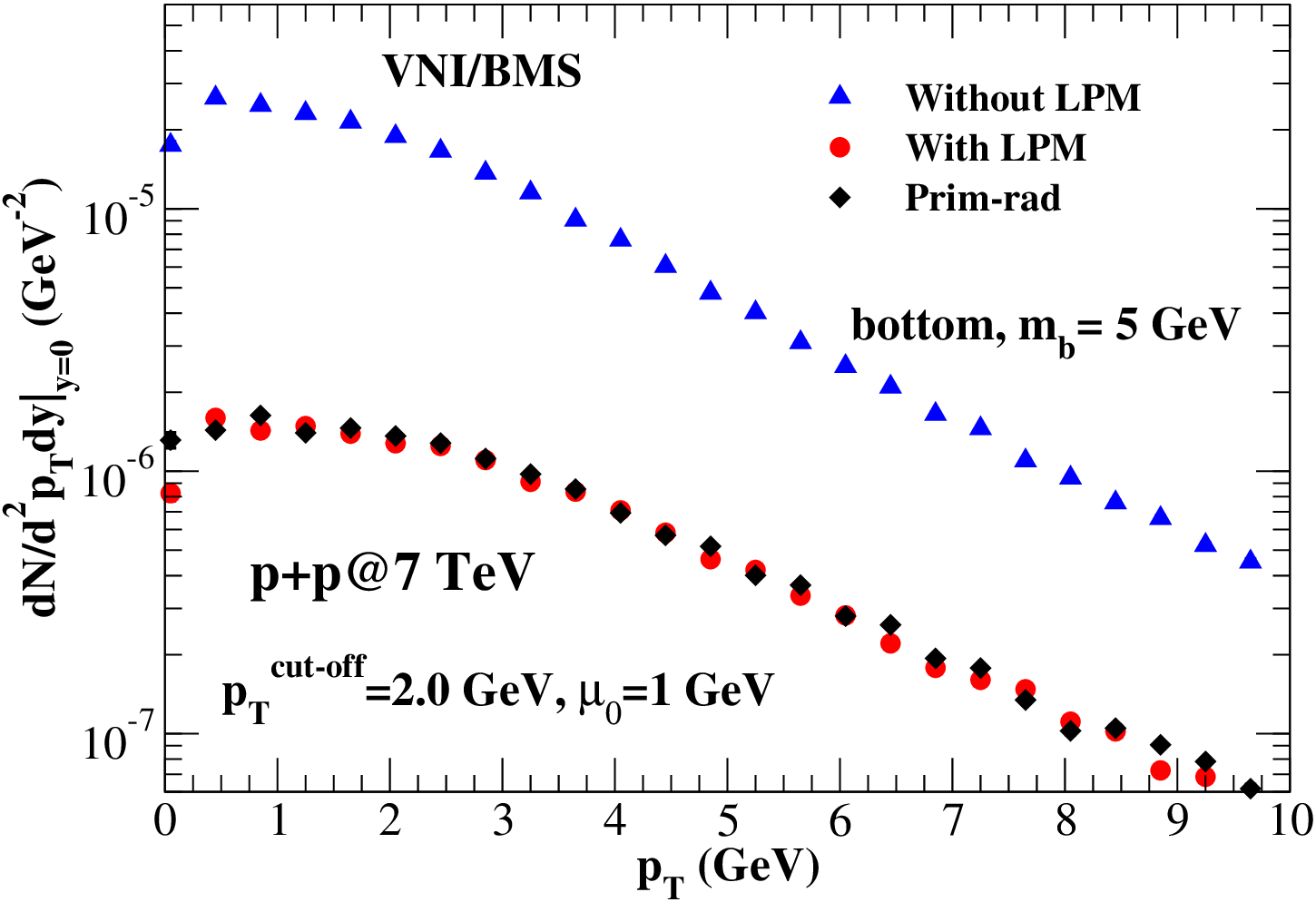}}
\caption{(Color online) The transverse momentum distribution of bottom quarks
produced in $pp$ collisions at $\sqrt{s}=$ 7  TeV at $y=0$ for $m_b=$ 4.00 GeV (upper
panel), 4.75 GeV (middle panel) and 5.00 GeV (lower panel). 
The three calculations involve multiple collisions among partons by neglecting and including the
LPM effect and collisions only among primary partons with radiations off the scattered partons.
}  
\label{b_pt_7}
\end{figure}

\begin{figure}[h]
\centerline{\includegraphics*[width=7.6 cm]{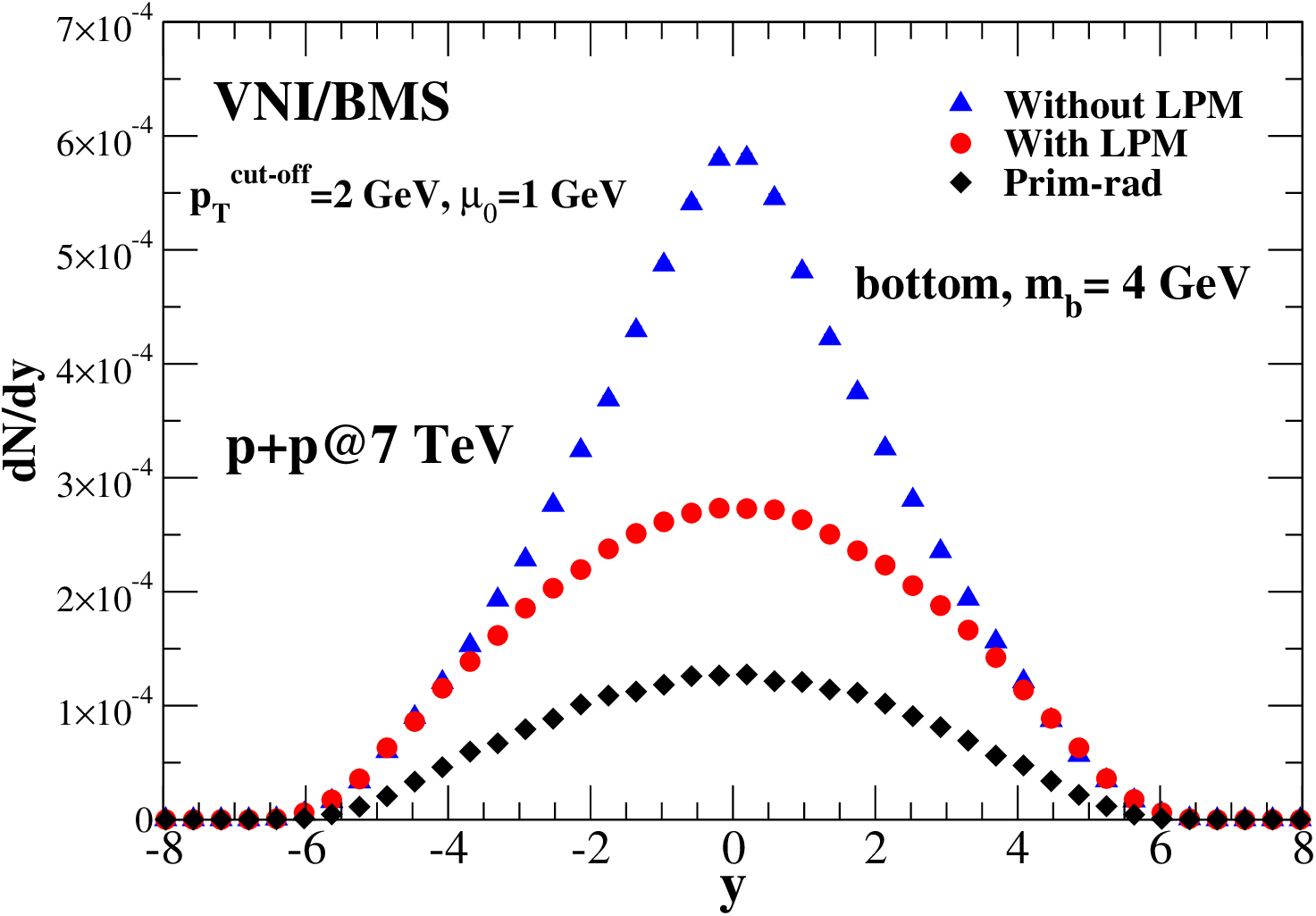}}
\centerline{\includegraphics*[width=7.6 cm]{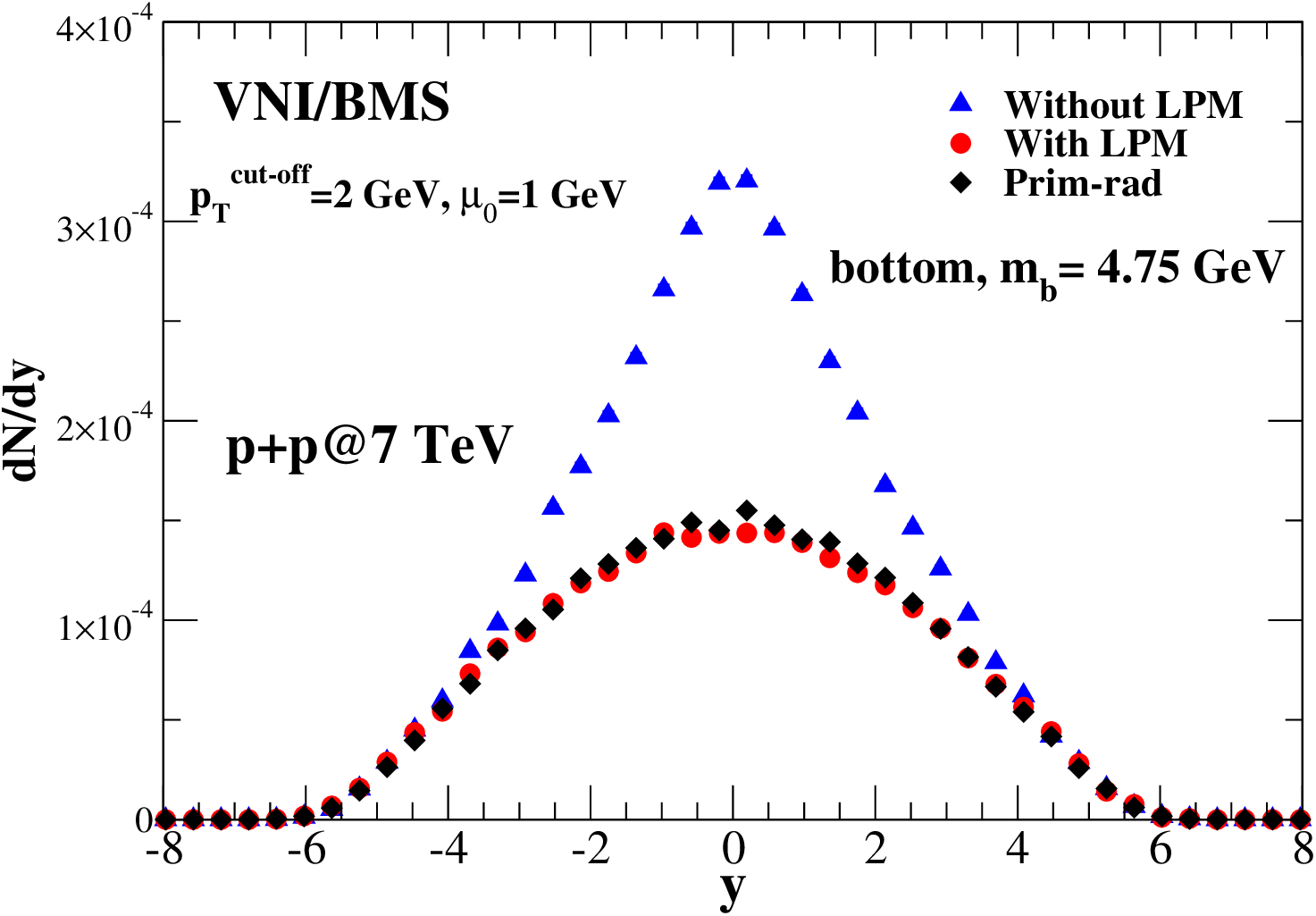}}
\centerline{\includegraphics*[width=7.6 cm]{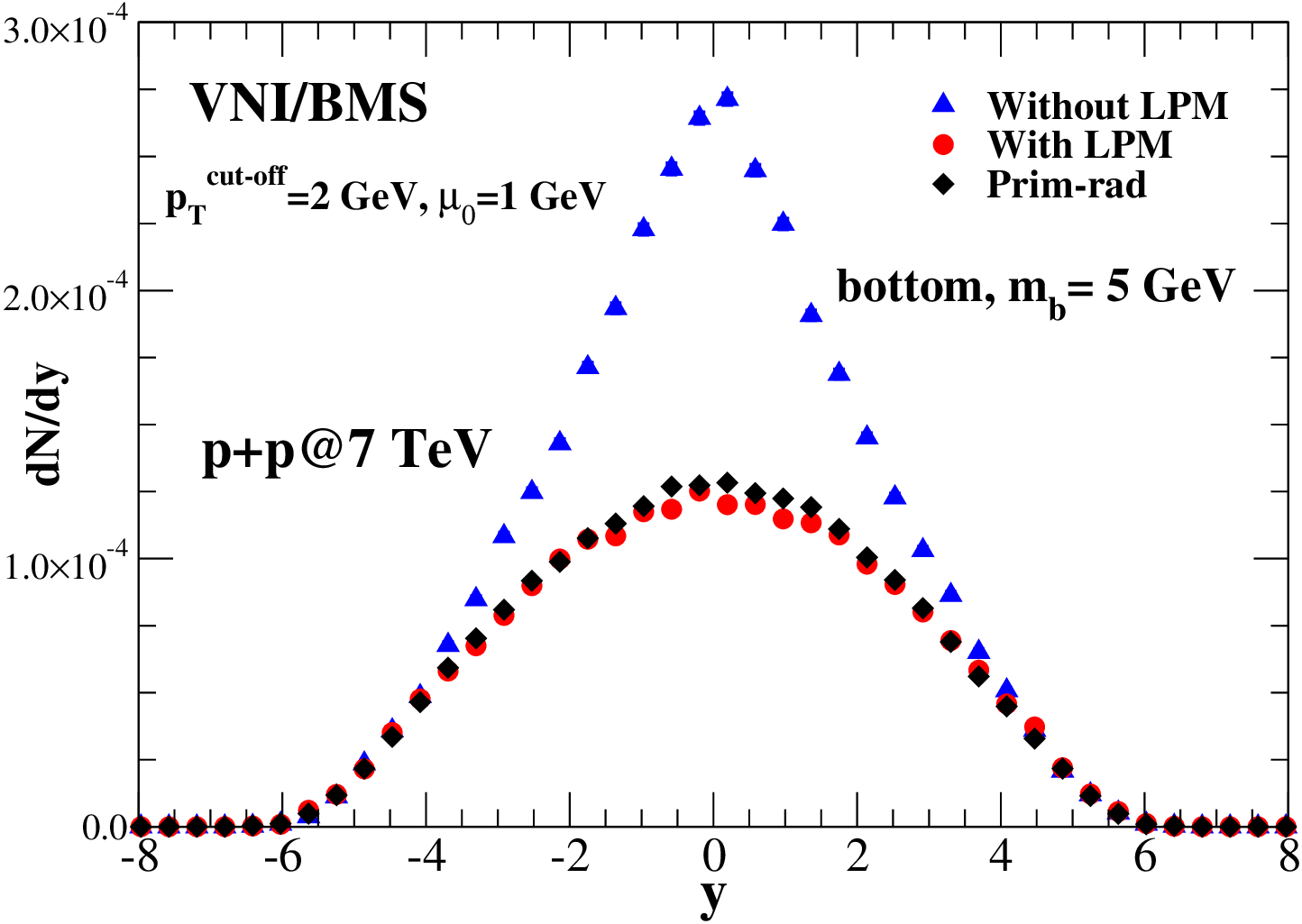}}
\caption{(Color online) The $p_T$ integrated rapidity spectra for bottom quarks
produced in $pp$ collisions at $\sqrt{s}=$ 7  TeV for $m_b=$ 4.00 GeV (upper
panel), 4.75 GeV (middle panel) and 5.00 GeV (lower panel). The 
three calculations involve multiple collisions among partons by neglecting and including the
LPM effect and collisions only among primary partons with radiations off the scattered partons.
}  
\label{b_y_7}
\end{figure}

\section{Formulation}

\begin{figure}[h]
\centerline{\includegraphics*[width=7.6 cm]{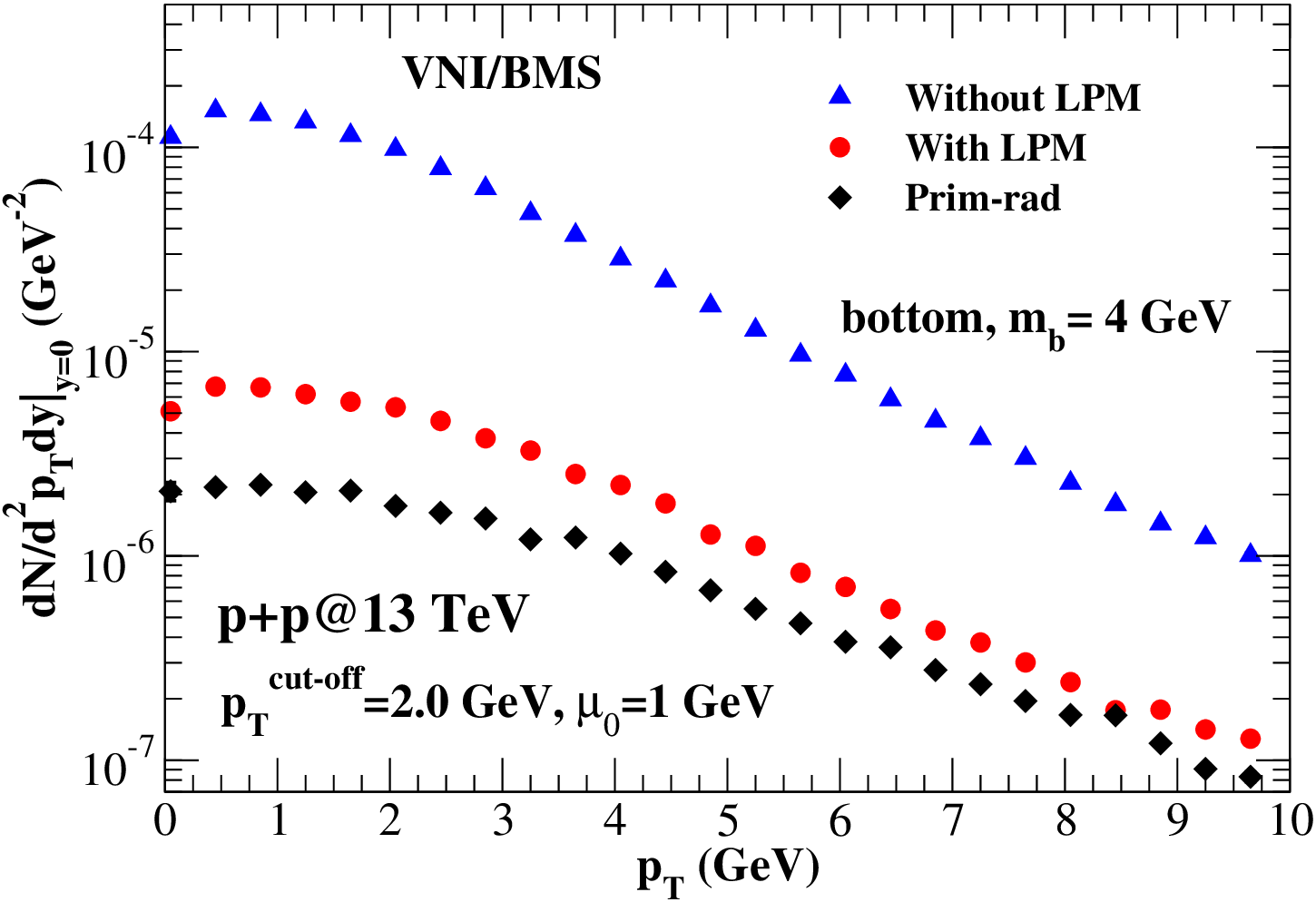}}
\centerline{\includegraphics*[width=7.6 cm]{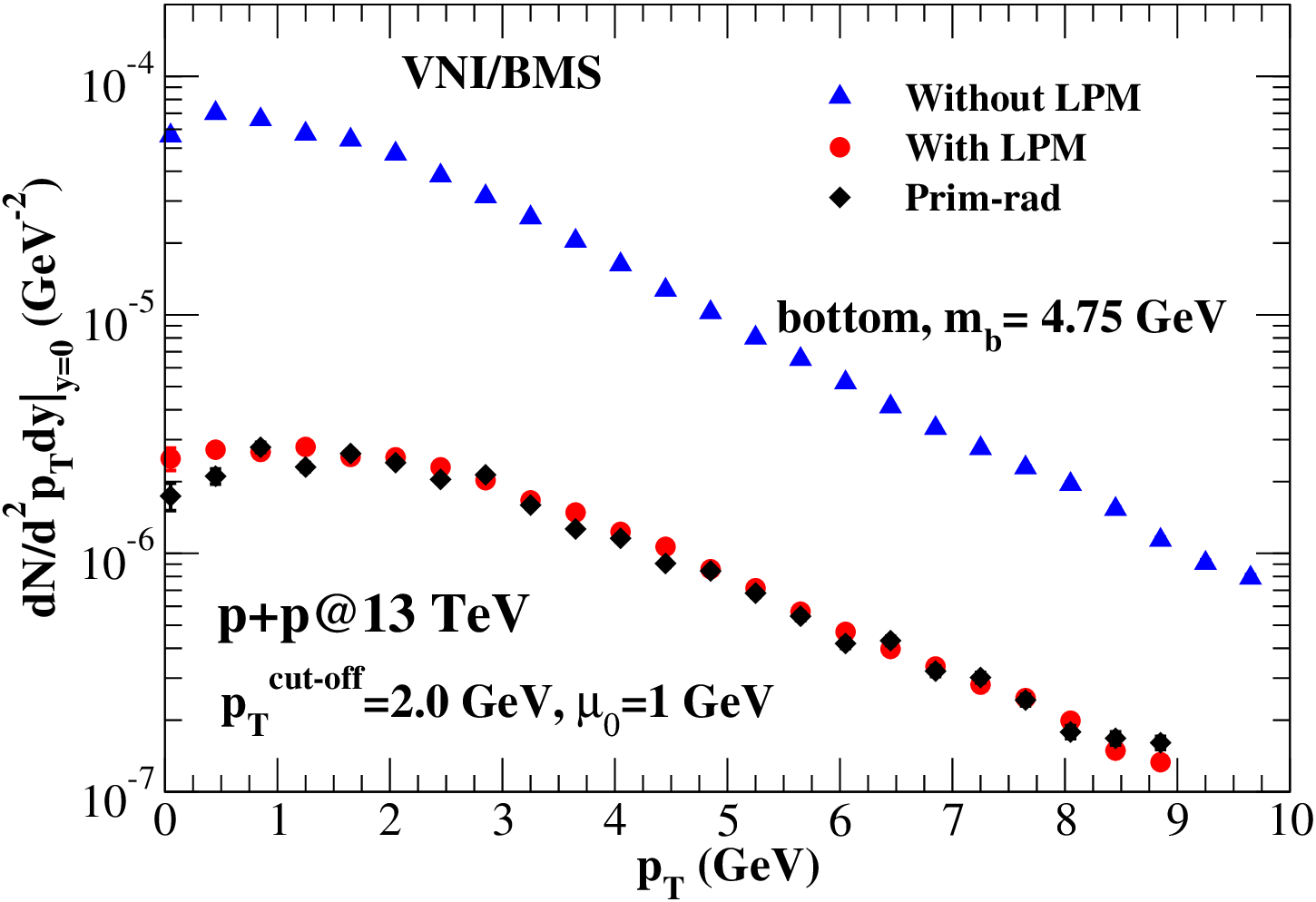}}
\centerline{\includegraphics*[width=7.6 cm]{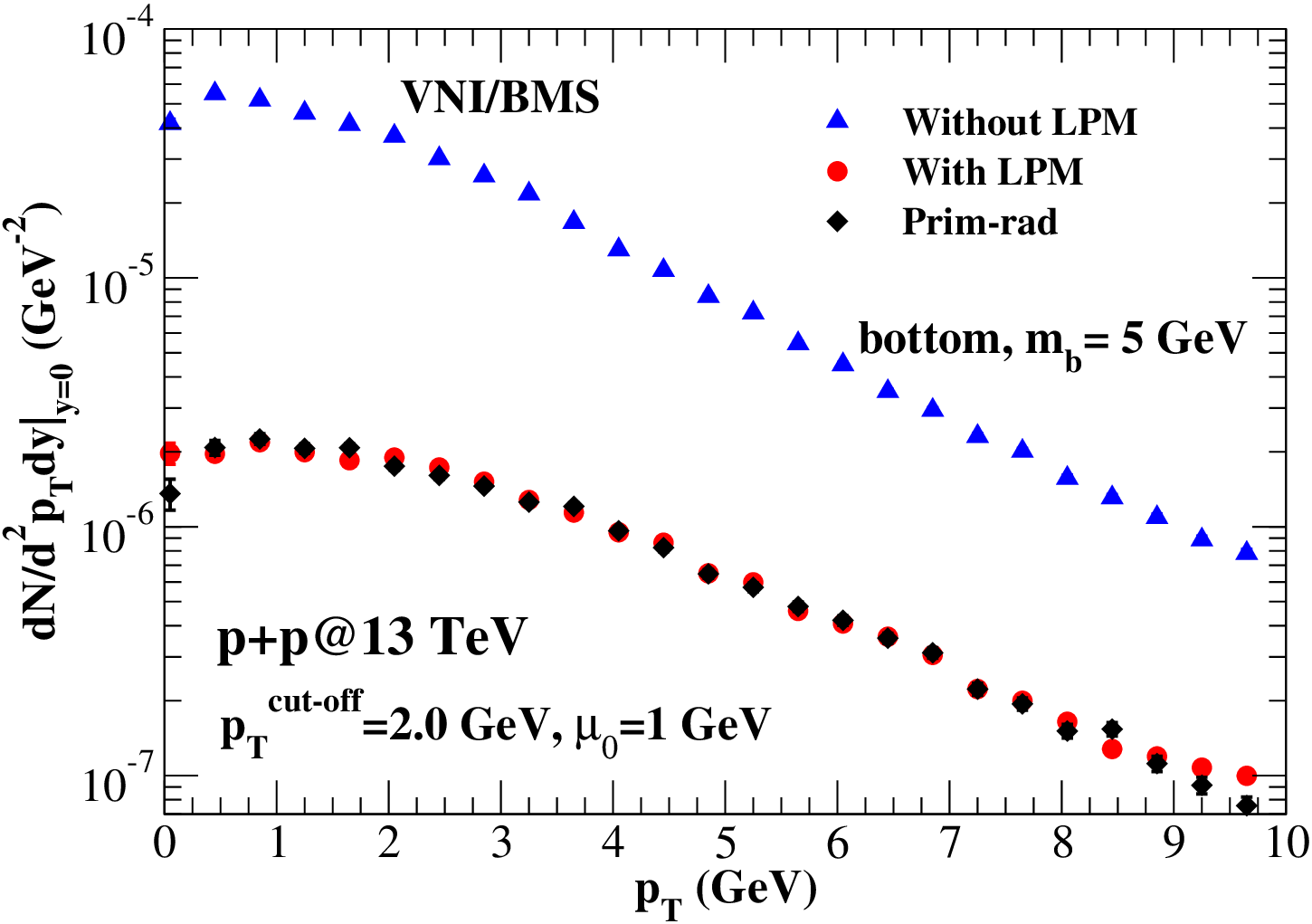}}
\caption{(Color online) The transverse momentum distribution of bottom quarks
produced in $pp$ collisions at $\sqrt{s}=$ 13 TeV at $y=0$ for $m_b=$ 4.00 GeV (upper
panel), 4.75 GeV (middle panel) and 5.00 GeV (lower panel). The 
three calculations involve multiple collisions among partons by neglecting and including the
LPM effect and collisions only among primary partons with radiations off the scattered partons.
}  
\label{b_pt_13}
\end{figure}

\begin{figure}[h]
\centerline{\includegraphics*[width=7.6 cm]{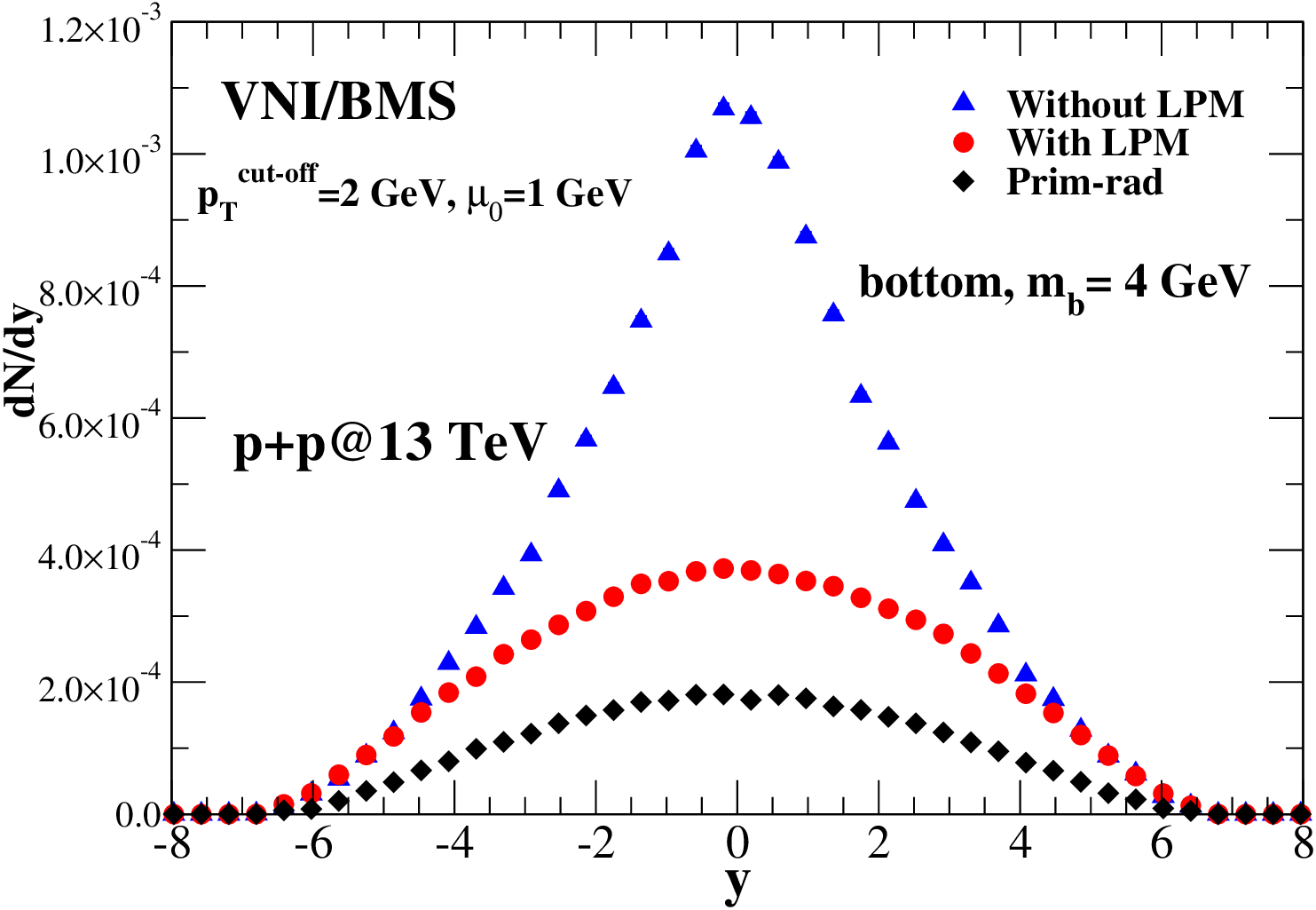}}
\centerline{\includegraphics*[width=7.6 cm]{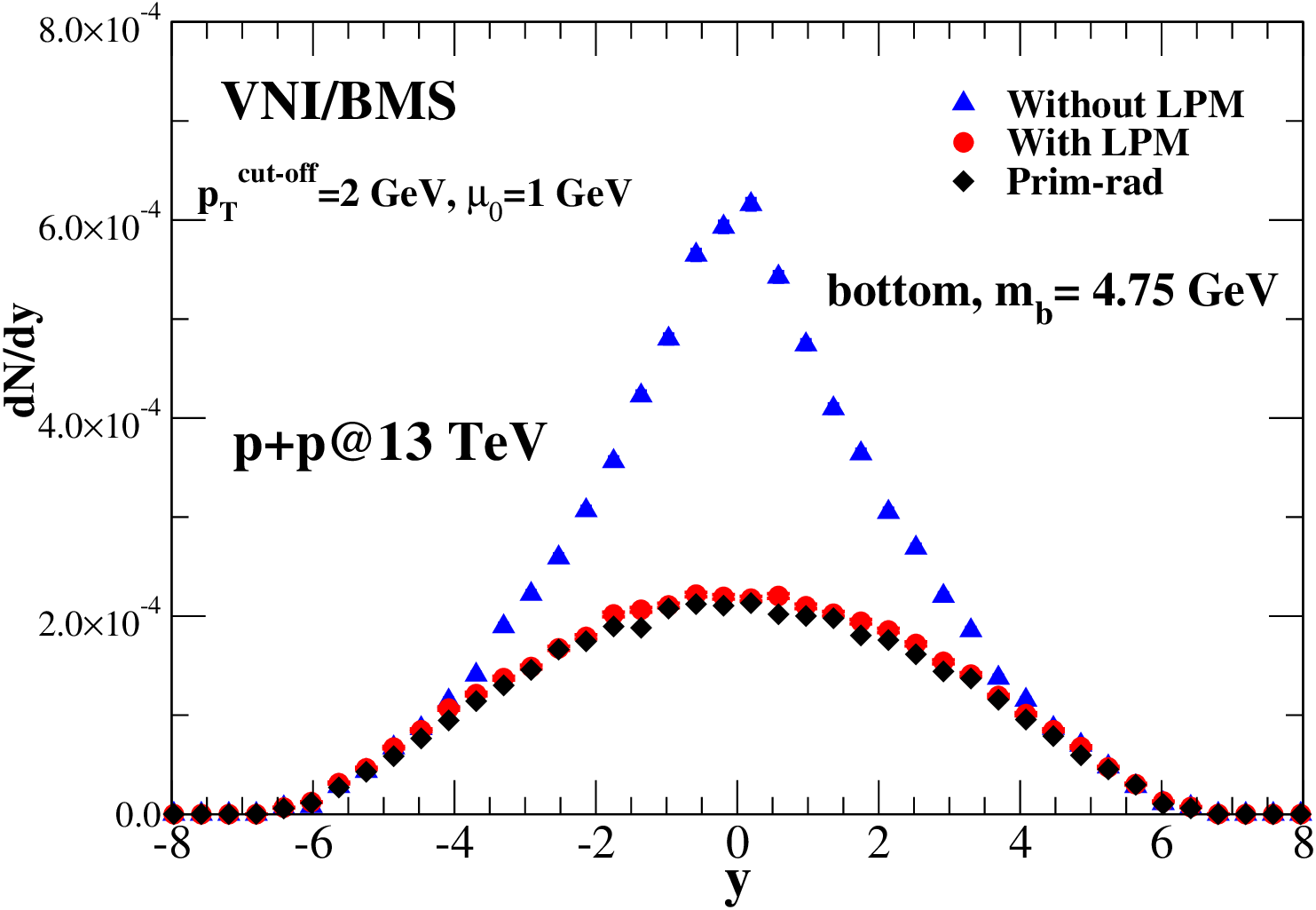}}
\centerline{\includegraphics*[width=7.6 cm]{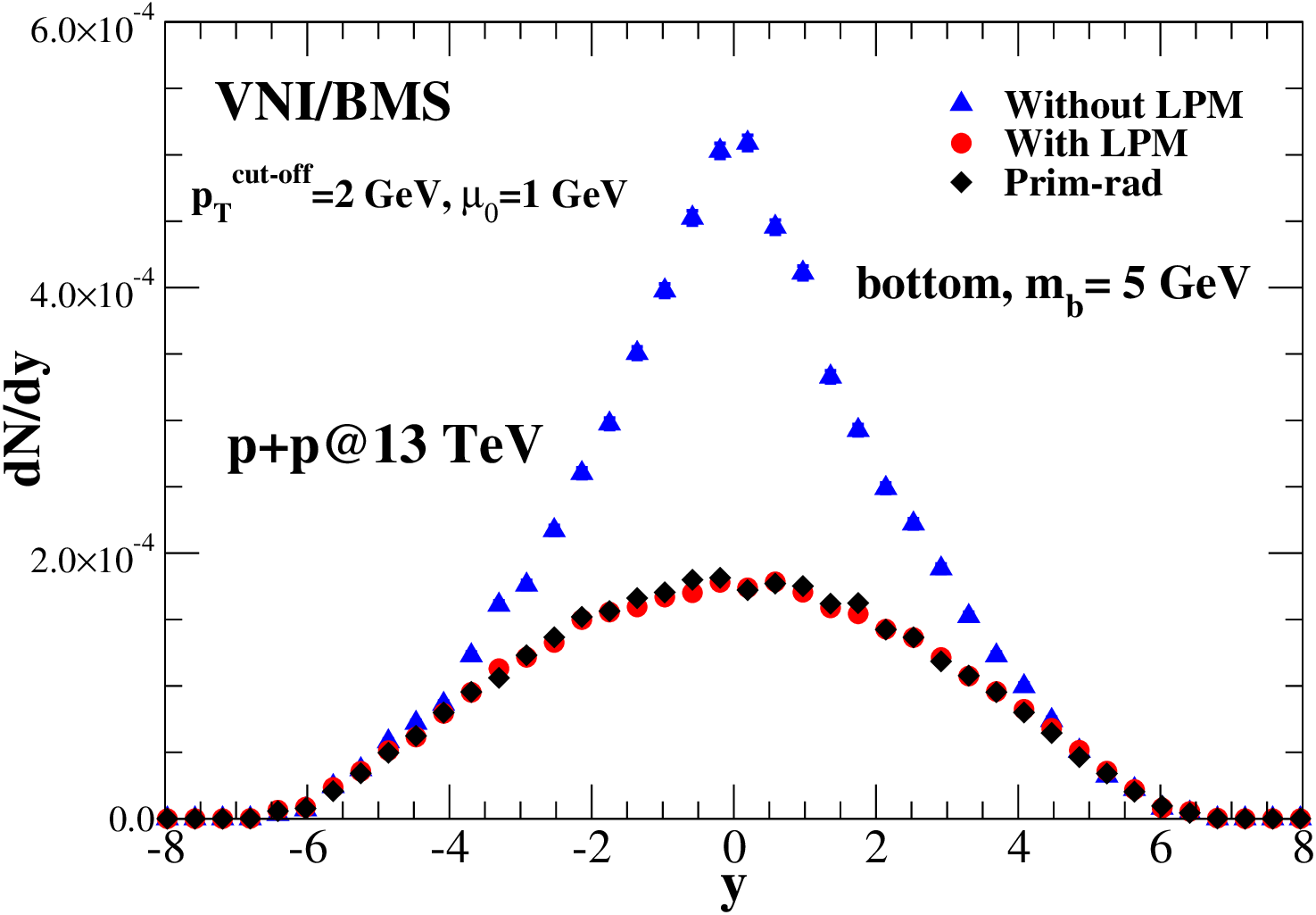}}
\caption{(Color online) The $p_T$ integrated rapidity spectra for bottom quarks
produced in $pp$ collisions at $\sqrt{s}=$ 7  TeV for $m_b=$ 4.00 GeV (upper
panel), 4.75 GeV (middle panel) and 5.00 GeV (lower panel). The 
three calculations involve multiple collisions among partons by neglecting and including the
LPM effect and collisions only among primary partons with radiations off the scattered partons.
}  
\label{b_y_13}
\end{figure}

We shall use the Monte Carlo implementation {\tt VNI/BMS} of the parton cascade model, discussed in
detail by several authors including the implementation of the LPM effect and heavy quark 
production~\cite{Geiger:1991nj,Bass:2002fh,Srivastava:2017bcm,Renk:2005yg,Bass:2002vm,Bass:2007hy}.
We obtain the time-evolution of the ensemble of quarks and gluons populating the nucleons 
(which populate the nuclei in the case of $AA$ collisions) on the basis of the Boltzmann equation. 
The $2\rightarrow 2$ scatterings between light quarks, heavy quarks and gluons, 
and the $2\rightarrow 3$ reactions via time-like branchings of the final-state
partons (see Ref.~\cite{Bass:2002fh,Altarelli:1977zs}) are included following the procedure adopted in 
{\tt PYTHIA}~\cite{Sjostrand:2006za}. We add that this procedure is known to account for higher order
effects in the parton scattering within Leading Logarithmic Approximation.

The two body matrix elements are regularized by implementing a $p_T^\text{cut-off}$. The
build up of soft gluons is regularized by implementing a virtuality cut-off $\mu_0$ for fragmentaions. 
We shall use a value of 2 GeV for 
$p_T^\text{cut-off}$ and keep $\mu_0$ fixed at 1 GeV. The LPM~\cite{Landau:1953gr} effect is implemented using the
procedure discussed in Ref.~\cite{Renk:2005yg} by assigning a formation time 
\begin{equation}
\tau = \frac{\omega}{k_T^2},
\end{equation}
where $k_T$ is its transverse momentum with respect
to the emitter and $\omega$ is its energy. We further require that the radiated particle does not
interact with other partons during the formation time. The radiating parton, however, is allowed
to interact, and if that happens during the formation time, the radiated particle is removed 
from the list of partons forming the system. It has been reported~\cite{Renk:2005yg,Srivastava:2018nfu}
that the dependence of the results obtained in the parton cascade model on $\mu_0$ is rather modest
once the LPM effect is accounted for.

The results for bottom production in pQCD depend on the mass for the bottom quark used in the matrix elements.
Values ranging from 4 to 5 GeV have been used in the literature. 
We give our results for the two extremes as well as for the one used most often, 
4.75 GeV~\cite{Cacciari:2012ny}.

The calculations are presented for $pp$ collisions at $\sqrt{s}=$ 7  and 13 TeV.

\begin{figure}[h]
\centerline{\includegraphics*[width=7.6 cm]{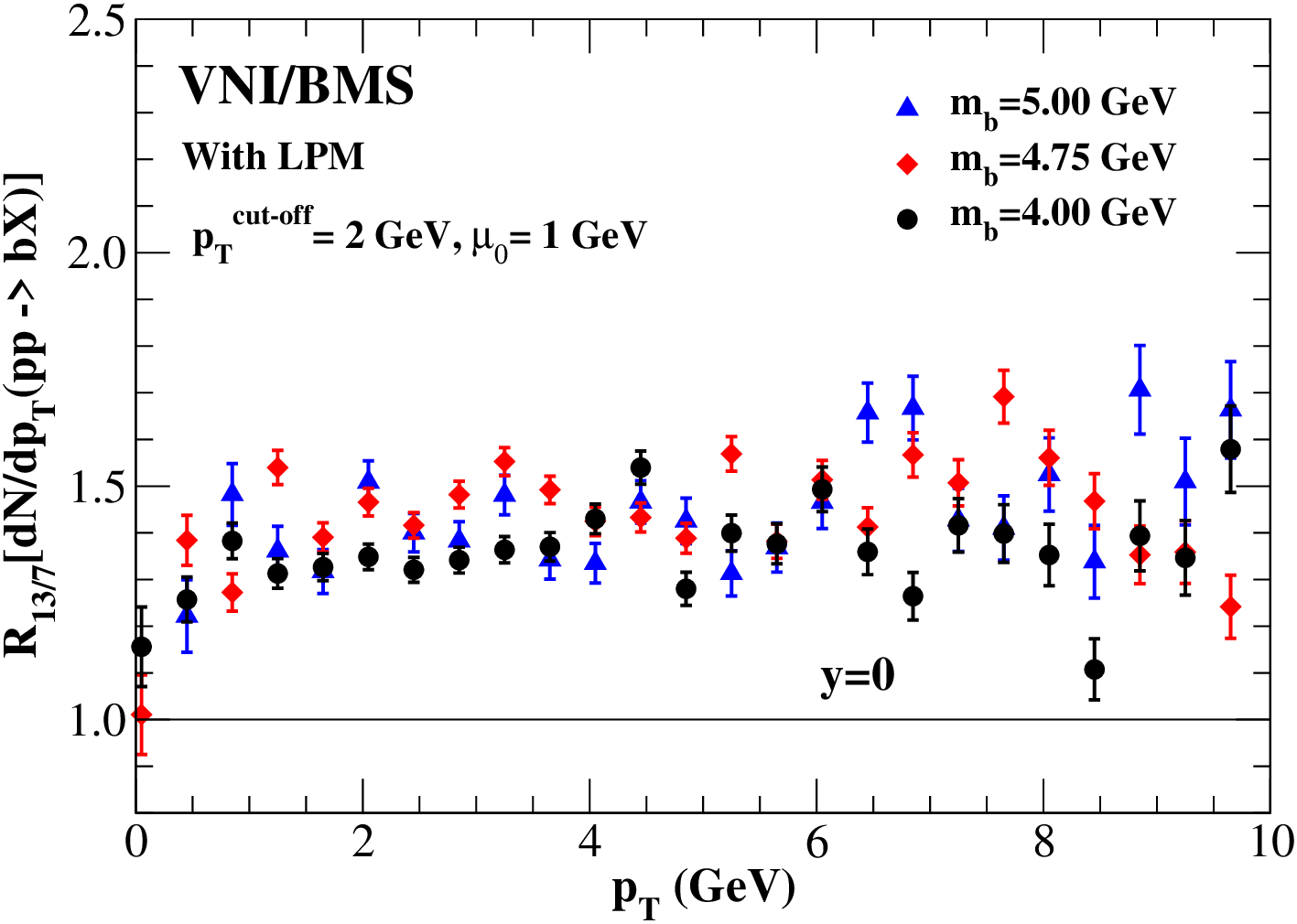}}
\centerline{\includegraphics*[width=7.6 cm]{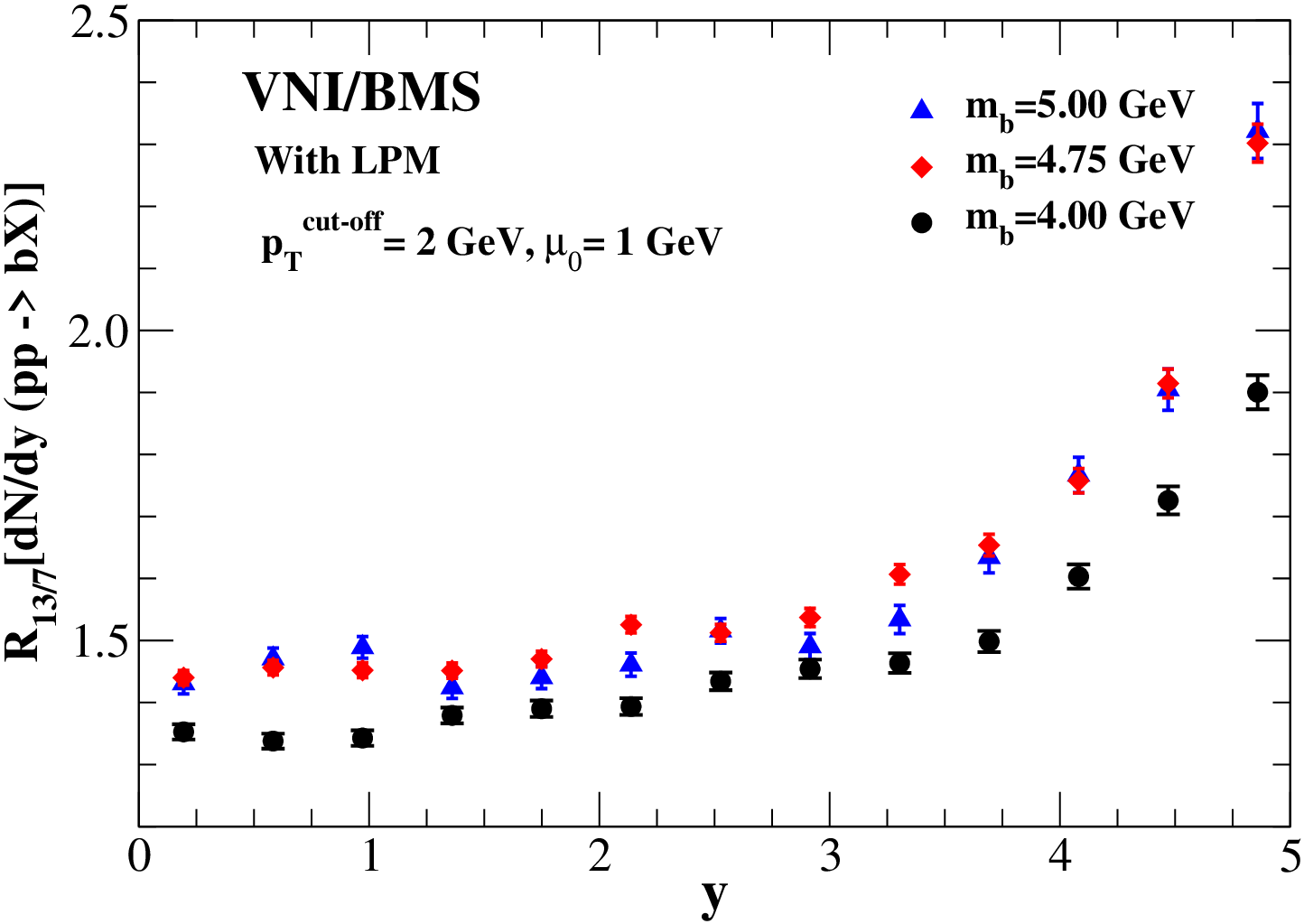}}

\caption{(Color online) Ratios of the transverse momentum spectra at $y=0$ (upper panel)
and $p_T$ integrated rapidity spectra (lower panel) for bottom quarks at
at $\sqrt{s}$= 13 and 7  TeV, for $m_b=$ 4.00, 4.75 and 5.00 GeV.} 
\label{ratio}
\end{figure}

\section{Results}

We discuss our results for three sets of calculations. In the first case, we put the formation time
of the gluons radiated off a final state parton in a hard scattering as zero, which corresponds to ignoring the
consequences of the
LPM effect.These results are given only as a reference as we do know that LPM effect definitely affects the dynamics
of light partons at low transverse momenta rather substantially.

Next we give results where only primary partons (the partons which initially populate the nucleons)
interact and radiate after scattering and thus exclude multiple scatterings of partons. We emphasize that by multiple scattering we imply a scenario where a given parton undergoes several collisions.
 
Finally we give our results for calculations accounting 
for multiple scatterings as well as the LPM effect. If the results for the later two calculations differ significantly,
it would indicate that the bottom quark production is affected by multiple scatterings in realistic situations after
LPM effect is accounted for. We recall that there are indications~\cite{Srivastava:2018nfu} that the results for charm production {\it are} quite sensitive to these differing scenarios 
for lower transverse momenta and central rapidities.  

In Fig.~\ref{b_pt_7} we give our results for the transverse momentum distribution of bottom quarks for the three cases
mentioned above and for the three masses.

We see that for all the cases the complete neglect of the LPM effect increases the production of bottom quarks considerably
as compared to the case when the LPM effect is accounted for. This, we think happens as the formation time
of the gluons delays their materialization, while the system continues to expand and dilute, thus reducing the
chances of their undergoing multiple scatterings. In any case mutiple scatterings which are hard enough to
 produce a pair of $b\overline{b}$ quarks are rather rare.

We also see that once LPM effect is accounted for, the results for bottom production with and without inclusion of
 multiple scatterings are nearly identical for the more realistic values of $m_b$, viz. 4.75 or 5.00 GeV.
These aspects become even more clear when we look at the $p_T$ integrated rapidity spectra (Fig~\ref{b_y_7}), 
where we see that
for these masses the rapidity spectra for bottom quarks are quite close when the LPM effect is accounted for. 
This suggests that
the inclusion of multiple scatterings does not lead to additional production of bottom quarks 
for reasonable values of $m_b$.
The deviations seen for $m_b=$ 4.00 GeV are similar in nature,
 though much smaller than that for the production of charm quarks
reported earlier~\cite{Srivastava:2018nfu}

The corresponding results for $pp$ collisions at 13 TeV are given in Figs.~\ref{b_pt_13} and \ref{b_y_13}.
We see that, while there is a larger production of bottom quarks as the energy rises, the relative trends of the 
productions for different masses and the three production scenarios remain similar to the case of collisions at 
7  TeV earlier.

These results suggest that when LPM effect is accounted for the contribution of multiple scatterings
of partons (where the same parton interacts repeatedly) to bottom production is rather marginal
for reasonable values of the $m_b$. 

Finally, we give the ratio of production of bottom quarks as a function of transverse momenta for $y=0$ 
and as a function of rapidity for $p_T$ integrated results (Fig.~\ref{ratio}). We see that the production
of bottom quarks at 13 TeV is about 1.4 times large than that at 7  TeV as a function of $p_T$ at
$y=0$, while the $p_T$ integrated results for the rapidity spectra are close to a value of 1.5 at 
central rapidities and rise 
slowly to about 2 at more forward (backward) rapidities. We note that there is only a 
modest dependence of these ratios
on the mass of the bottom quarks used in the calculations.

 We add that these corresponding ratios for
cross-sections can be obtained by multiplying these 
results with $\sigma_{\text{in}}(7 \ {\rm TeV})/\sigma_{\text{in}}(13 \ {\rm TeV}) \approx 0.9$, and are similar to
the results obtained from Fixed Order + Next to Leading Log (FONLL) calculations 
reported earlier~\cite{Cacciari:2012ny}.

\section{Summary and Conclusions}
We have reported results of calculations for production of bottom quarks in $pp$ collisions at 
Large Hadron Collider energies using Parton Cascade Model. 

We find that for reasonable values
of the mass of bottom quark ($\approx$ 4.75 GeV or more) multiple scattering of partons does
not play any significant role for their production, once Landau Pomeranchuk Midgal effect is switched on. This,
we feel, happens as multiple scatterings strongly affect the dynamics of partons at lower transverse
momenta, which however may not be able to produce bottom quarks which have a large mass, in contrast to the
case of charm quarks reported earlier.

This suggests that the usual nuclear modification factor $R_{\text AA}$ for bottom quarks
can be used with relative confidence to obtain medium modification of bottom production in
relativistic collision of heavy nuclei at Large Hadron Collider. 

These results taken along with our earlier findings about charm quarks~\cite{Srivastava:2018nfu}
elevate bottom quarks to a pre-eminent position to study medium modification of heavy quark
production in $AA$ collisions by comparing the same to that for $pp$ collisions, while the charm 
quarks are elevated to a pre-eminent position for confirming the advent of an interacting medium
in $pp$ collisions at LHC energies, where multiple semi-hard partonic collisions take place even when
LPM effects are accounted for.

\section*{Acknowledgments} 
DKS gratefully acknowledges the support by the Department of Atomic
Energy. We thank the Computer and Informatics Group,  Variable 
Energy Cyclotron Centre, Kolkata for providing the computing facility
necessary for these studies.


\end{document}